\documentclass[aps,prl,reprint,superscriptaddress]{revtex4-2}
\usepackage{graphicx}
\usepackage{dcolumn}
\usepackage{bm}
\usepackage{romannum}
\usepackage[utf8]{inputenc}
\usepackage[T1]{fontenc}
\usepackage{booktabs, array, float, tabularx, booktabs, lipsum, multirow, amstext}
\usepackage{amsmath}
\usepackage{amsfonts}
\usepackage{mathptmx}
\usepackage{mathrsfs}
\usepackage{amssymb}
\DeclareMathAlphabet{\mathcal}{OMS}{cmsy}{m}{n}
\usepackage{siunitx}
\usepackage[svgnames,x11names,table]{xcolor}
\usepackage[version=4]{mhchem}
\usepackage{appendix}
\usepackage{url}
\usepackage{hyperref}

\hypersetup{
	colorlinks=true,
	linkcolor=Blue,
	filecolor=Blue,
	citecolor=Blue,      
	urlcolor=Blue,
}

\usepackage{caption}    
\usepackage{subcaption} 
\usepackage{microtype}
\usepackage{ragged2e}
\usepackage{orcidlink}
\usepackage{float}

\begin{document}

\title{Probing false vacuum decay and bubble nucleation in a Rydberg atom array}

\author{Yu-Xin Chao}
\thanks{These authors contributed equally to this work.}
\affiliation{State Key Laboratory of Low Dimensional Quantum Physics, Department of Physics, Tsinghua University, Beijing 100084, China}
\affiliation{Beijing Academy of Quantum Information Sciences, Beijing 100193, China}

\author{Peiyun Ge}
\thanks{These authors contributed equally to this work.}
\affiliation{State Key Laboratory of Low Dimensional Quantum Physics, Department of Physics, Tsinghua University, Beijing 100084, China}
\affiliation{Beijing Academy of Quantum Information Sciences, Beijing 100193, China}

\author{Zhen-Xing Hua}
\affiliation{State Key Laboratory of Low Dimensional Quantum Physics, Department of Physics, Tsinghua University, Beijing 100084, China}

\author{Chen Jia}
\affiliation{State Key Laboratory of Low Dimensional Quantum Physics, Department of Physics, Tsinghua University, Beijing 100084, China}

\author{Xiao Wang\,\orcidlink{0000-0003-2898-3355}}
\affiliation{Department of Physics, Cornell University, Ithaca, NY, USA}

\author{Xinhui Liang}
\affiliation{State Key Laboratory of Low Dimensional Quantum Physics, Department of Physics, Tsinghua University, Beijing 100084, China}

\author{Zongpei Yue}
\affiliation{State Key Laboratory of Low Dimensional Quantum Physics, Department of Physics, Tsinghua University, Beijing 100084, China}

\author{Rong Lu}
\affiliation{State Key Laboratory of Low Dimensional Quantum Physics, Department of Physics, Tsinghua University, Beijing 100084, China}

\author{Meng Khoon Tey\orcidlink{0000-0002-4861-527X}}
\email{mengkhoon\_tey@mail.tsinghua.edu.cn}
\affiliation{State Key Laboratory of Low Dimensional Quantum Physics, Department of Physics, Tsinghua University, Beijing 100084, China}
\affiliation{Frontier Science Center for Quantum Information, Beijing 100084, China}
\affiliation{Hefei National Laboratory, Hefei, Anhui 230088, China}

\author{Xiao Wang\,\orcidlink{0000-0002-3022-7260}}
\email{xw970921@gmail.com}
\affiliation{State Key Laboratory of Low Dimensional Quantum Physics, Department of Physics, Tsinghua University, Beijing 100084, China}
\affiliation{Beijing Academy of Quantum Information Sciences, Beijing 100193, China}
\affiliation{Department of Physics, University of Oxford, Oxfordshire, UK}

\author{Li You\orcidlink{0000-0002-3919-4768}}
\email{lyou@tsinghua.edu.cn}
\affiliation{State Key Laboratory of Low Dimensional Quantum Physics, Department of Physics, Tsinghua University, Beijing 100084, China}
\affiliation{Beijing Academy of Quantum Information Sciences, Beijing 100193, China}
\affiliation{Frontier Science Center for Quantum Information, Beijing 100084, China}
\affiliation{Hefei National Laboratory, Hefei, Anhui 230088, China}

\date{\today}

\begin{abstract}

In quantum field theory (QFT), the ``vacuum'' is not just empty space but the lowest-energy state of a quantum field. If the energy landscape has multiple local minima, the local ground states are the false vacuum (FV) which can tunnel towards the global ground state (true vacuum, TV). This process exhibits signature akin to classical supercooled gas transitions and many-body tunneling in discrete quantum systems. Here, we study the FV decay and bubble nucleation in a Rydberg atom ring. The $1/r^6$ van-der-Waals interactions and individual-site addressability allow us to explore physics beyond the standard Ising model. We observe that the FV decay rate decreases exponentially with the inverse of the symmetry-breaking field, directly mirroring QFT predictions. Moreover, we demonstrate that even minor deviations from the ideal metastable state can cause a stark departure from this universal scaling law. Extending beyond short-time decay dynamics, we also examine resonant bubble nucleation, a feature distinctive to systems with discrete energy spectra. Our findings and methods open avenues for future studies of many-body tunneling in higher dimensions or more complex geometries.

\end{abstract}

\maketitle

{\it Introduction. ---} Quantum tunneling in many-body systems~\cite{2008_Bloch_ManyBodyPhysics,2015_JiangYing_Water,2016_Anders_DirectQuantumTunneling,2016_Boixo_MultiQubitTunnelling,2025_YinChao_MetastableState} exhibits dynamics intrinsically richer than those of single particles or collective phases described by a single order parameter. A paradigmatic example is false vacuum decay (FVD) in quantum field theory (QFT). This process involves the tunneling of a metastable vacuum towards a lower-energy state, which is typically described by instanton solutions~\cite{1977_Coleman_Instanton,1977_Callan_1stQuantumCorrection,1990_Farhi_CreatUniverse,2018_Brown_ThinWall,2022_Devoto_FVDReview}. A distinct feature of such tunneling is the nucleation and growth of ``bubbles''---locally transformed regions---within the metastable state~\cite{Langer1967CondensationPoint,2021_Lagnese_FVDinSpinChains,2021_Sinha_QuantizedBubbleNucleation,2024_Darbha_AFM}. Such decay processes are ubiquitous across diverse physical systems, spanning from classical to quantum, continuous to discrete, open to closed~\cite{1969_Langer_MetastableState,1990_RMP_ReactionRate,2001_Stillinger_Supercooled,2016_Lesanovsky_MetastableOpen}. Potential realizations have been explored in ultracold atoms~\cite{2017Fialko_BEC,Billam_2019,Ng_2021}, trapped ions~\cite{2013Hauke_trappedions,2016Yang_trappedions}, and superconducting circuits~\cite{2021Abel_superconducting}. Recent experimental progress includes observation of bubble emergence~\cite{2024_Zenesini_SuperFluid, 2025_Monroe_irontrap}, bubble hopping and interaction dynamics in weak coupling~\cite{2025_Vodeb_DWave}, and excitation modes via Fourier spectroscopy of quench dynamics~\cite{2024_ZhenShengYuan_OpticalLattice}.

Rydberg atom arrays have recently emerged as a powerful and versatile platform for simulating quantum many-body dynamics~\cite{2010_Saffman_RMP, 2022_Saffman_QuantumComputer, 2021_Lukin_256atom, 2023_ChenCheng_CSB, 2025_KKN_CriticalPhenomena}. These systems, featuring individually trapped and detected atoms with tunable interactions between Rydberg states, can be utilized to simulate diverse spin models~\cite{2025_Xinhui_OTOC, 2025_Yue_Disorder, 2025_QiaoMu_TJmodel}. Furthermore, the ability to individually address and manipulate atoms with lasers enables the exploration of even richer quantum phases and non-equilibrium phenomena~\cite{2024_Browaeys_LocalControl, 2025_XuPeng_FiberArray, 2025_LinLi_LatticeGaugeTheory}.

In this work, we simulate FVD in an antiferromagnetic (AFM) Ising ring using a programmable Rydberg atom array. By applying a site-dependent longitudinal field, we lift the $\mathbb{Z}_2$ degeneracy of the two Néel ground states, casting one as a false vacuum (FV) and the other as the true vacuum (TV). We first probe the short-time dynamics of a Néel state after a quench, observing the characteristic exponential decay of AFM order. We then study the dynamics of a more refined metastable state, the pre-quench ground (PQG) state. We show that the PQG decay rate is exponentially suppressed by the inverse symmetry-breaking field, a feature that persists down to the weak-field regime and contrasts sharply with the Néel state. This behavior holds even for a generalized Ising model incorporating both global and staggered potentials, as well as $1/r^6$ van-der-Waals interactions. Finally, we explore the long-time dynamics to study resonant bubble nucleation, a process characteristic of systems with a discrete energy spectrum.

{\it Model and experimental setup. ---}
\begin{figure*}[!ht]
	\centering
    \includegraphics[width=1.9\columnwidth]{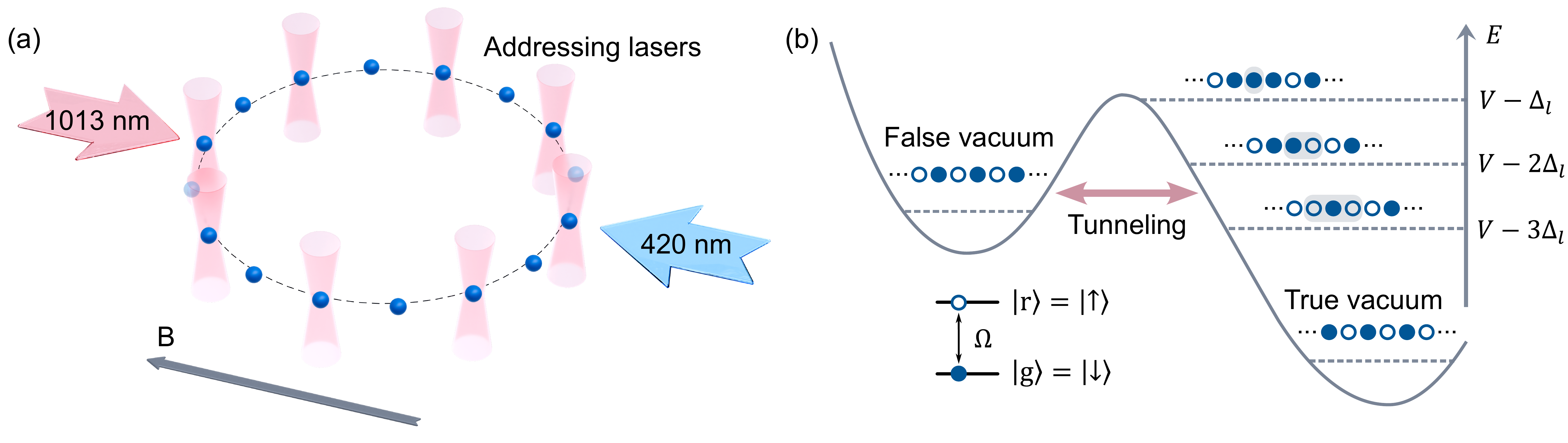}
	\caption{Simulating false vacuum decay with a programmable Rydberg atom array. (a) Schematic of the experimental setup. Neutral atoms are trapped in a ring geometry and illuminated globally by the 420-nm and 1013-nm lasers, coupling the ground state to a high-lying Rydberg state (70S). A set of far-detuned, site-selective addressing lasers (pink beams) illuminates every other atom, generating a staggered detuning. A magnetic field $B$ parallel to the 420-nm laser defines the quantization axis. (b) Energy landscape for false vacuum decay in an antiferromagnetic Ising model. The staggered longitudinal field breaks the degeneracy of the two N\'eel-ordered ground states, creating FV and TV states. Quantum tunneling from the FV towards the TV proceeds via the nucleation of TV bubbles (illustrated as gray domains of flipped spins). The dashed lines denote static energies ($\Omega=0$) of representative spin configurations along the unfolded ring.
    }
    \label{fig_setup}
\end{figure*}
As illustrated in Fig.~\ref{fig_setup}(a), our system is a ring-shaped array consisting of $N$ (with $N$ even) uniformly spaced $^{87}\rm{Rb}$ atoms, with the pseudo-spin-$\frac{1}{2}$ encoded in the ground state $|\!\downarrow\rangle \equiv |5S_{1/2}, F=2, m_F=2\rangle$ and the Rydberg state $|\!\uparrow\rangle \equiv |70S_{1/2}, m_J=1/2 \rangle$. The spin states are coupled via a two-photon transition (420-nm and 1013-nm), through the intermediate state $|6P_{3/2}, F=3, m_F=3\rangle$, with an effective Rabi frequency $\Omega$. The Hamiltonian governing the system is
\begin{equation} \label{eq_H}
    \hat{H}/ \hbar = \frac{\Omega}{2} \sum_{j=1}^N \hat{\sigma}^x_j + \sum_{j=1}^N \left[ -\Delta_g + (-1)^j\Delta_l \right] \hat{n}_j + \sum_{i<j} V_{i,j} \hat{n}_i \hat{n}_j,
\end{equation}
where $\hat{n}_j = |\!\uparrow \rangle_j\!\langle \uparrow\!|$ ($\hat{g}_j = |\!\downarrow \rangle_j \!\langle \downarrow\!|$) is the Rydberg (ground) state occupation operator, $\hat{\sigma}^x_j = |\!\uparrow\rangle_j\!\langle\downarrow\!| + \text{h.c.}$ is the spin-flip operator. The term $V_{i,j} \propto |i-j|^{-6}$ describes the van-der-Waals interaction, which is antiferromagnetic ($V_{i,j}>0$) and dominated by the nearest-neighbor coupling $V \equiv V_{i,i+1}$. The global detuning $\Delta_g$ and staggered detuning $(-1)^j \Delta_l$ are engineered by adjusting the two-photon detuning of 420-nm and 1013-nm lasers, in conjunction with the AC Stark shifts of site-selective addressing laser beams [Fig.~\ref{fig_setup}(a)]. Retaining only the nearest neighbor interactions and setting $\Delta_g=V$, the antiferromagnetic Hamiltonian (\ref{eq_H}) maps directly to the well studied ferromagnetic quantum Ising model with both transverse and longitudinal fields: $\hat{H}_f/ \hbar = -\sum_{j=1}^N \left[ \hat{\sigma}^z_j \hat{\sigma}^z_{j+1}+h_x \hat{\sigma}^x_j + h_z \hat{\sigma}^z_j\right]$, with $h_x=2\Omega/V$ and $h_z=2\Delta_l/V$. Crucially, by setting $\Delta_{g} \neq V$, we can explore a generalized Ising model and physics beyond the standard Ising model with this platform.

In the absence of the staggered field ($\Delta_l = 0$) and in the limit $\Omega \to 0$, the Hamiltonian $\hat{H}$ possesses a $\mathbb{Z}_2$ symmetry and features two degenerate Néel-ordered ground states. The staggered field $\Delta_l$ acts as the symmetry-breaking field, lifting this degeneracy (Fig. 1(b)). For $\Delta_l > 0$, the state $|\text{Néel}\rangle = |\!\downarrow\uparrow\dots\downarrow\uparrow\rangle$ becomes the high-energy metastable state (the FV), while $|\text{Néel}'\rangle = |\!\uparrow\downarrow\dots\uparrow\downarrow\rangle$ becomes the low-energy TV~\footnote{In the presence of a transverse field $\Omega$, the classical Néel states are not exact eigenstates; they serve as approximations of the vacuum states.}.

The decay from the FV proceeds via the nucleation of TV bubbles, as sketched in Fig.~\ref{fig_setup}(b). The smallest bubble can form in two ways: (i) flipping an $|\!\uparrow \rangle$ spin into $|\!\downarrow\rangle$, acquiring an energy of $\Delta E_{\uparrow \to \downarrow} = \Delta_g - \Delta_l$; or (ii) flipping a $|\!\downarrow \rangle$ spin into $\!|\uparrow \rangle$, causing an energy penalty of $\Delta E_{\downarrow\to\uparrow} = 2V - \Delta_g - \Delta_l$. These energy costs become degenerate, $\Delta E = V - \Delta_l$, in the standard Ising limit $\Delta_g = V$.

{\it FV decay of the Néel state in a generalized Ising model ---}
We begin by studying FVD in the generalized Ising regime with $\Delta_{g}=0.8V$. We prepare the initial $|\text{Néel}\rangle$ state (the FV) by applying AC Stark shifts to every other atom, followed by a Landau-Zener sweep~\cite{1932_Landau} to excite the addressed atoms to the Rydberg state (see Supplemental Material (SM) Sec.~\Romannum{1}~\cite{SM}). The decay dynamics are initiated by a quench of $\Omega$, $\Delta_g$ and $\Delta_l$, after which we track the decay of the AFM order parameter,
\begin{equation} \label{eq_M_AFM}
    \langle \hat{M}_{\mathrm{AFM}}\rangle = \frac{1}{N}\sum_{j=1}^N (-1)^j \langle\hat{\sigma}_j^z\rangle.
\end{equation}
This parameter, which is 1 for the FV and -1 for the TV, is rescaled as in reference~\cite{2021_Lagnese_FVDinSpinChains} 
\begin{equation} \label{eq_M_AFM_rescaled}
    M_{\rm{AFM}}^{\rm{~res}}(t) = \frac{\langle \hat{M}_{\mathrm{AFM}}(t)\rangle + \langle \hat{M}_{\mathrm{AFM}}(0)\rangle}{2\langle \hat{M}_{\mathrm{AFM}}(0)\rangle}.
\end{equation}

\begin{figure}[!ht]
	\centering
	\includegraphics[width=0.86\columnwidth]{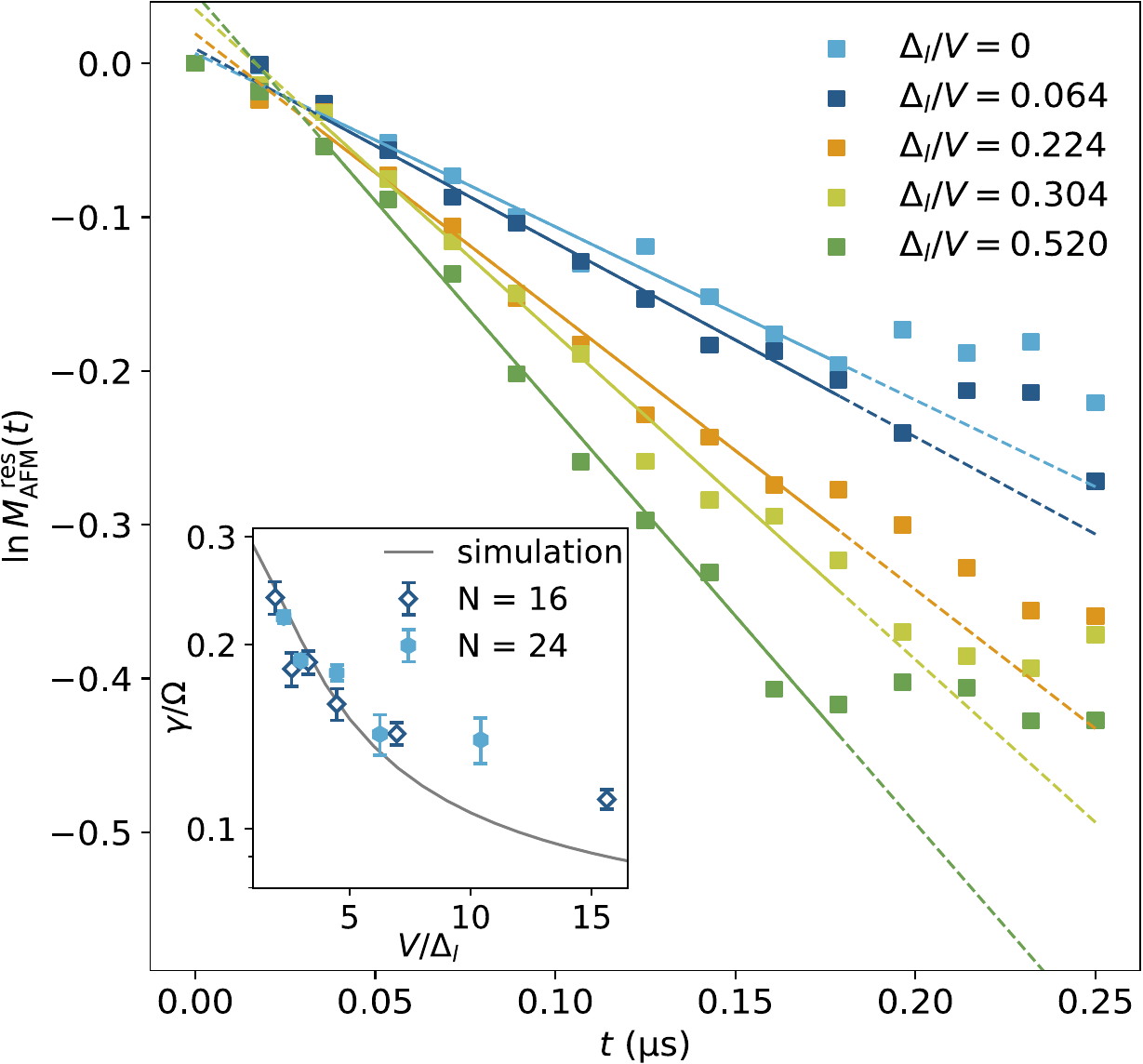}
	\caption{
    Dynamics and scaling of the false vacuum decay. Main panel: Measured time evolution of the rescaled antiferromagnetic (AFM) order for an $N=16$ atom ring, plotted on a logarithmic scale  for different $V/\Delta_l$ ratios. Experimental data (solid squares) are averaged over 300 realizations.
    Solid lines are exponential fits $M_{\text{AFM}}^{\text{~res}}(t) \propto e^{-\gamma t}$ to the data within time windows $t \in [0.02,0.18]~\rm{\mu}s$, used to extract the decay rate $\gamma$. Dashed lines are extensions of these fits.
    Inset: Extracted decay rate $\gamma$, normalized by $\Omega$, as a function of the inverse local staggered field $V/\Delta_l$. Experimental data for $N=16$ (empty diamonds) and $N=24$ (solid hexagons) are presented on a semi-logarithmic scale with statistical error bars. The gray line is the decay rate extracted from numerical simulations for $N=16$ without considering any experimental errors.
    }
    \label{fig_decay}
\end{figure}
 
The main panel of Fig.~\ref{fig_decay} presents the measured decay of $M_{\rm{AFM}}^{\rm{~res}}(t)$ for an $N=16$ ring at different $\Delta_l$, with fixed parameters $\Omega/2\pi=1.8$~MHz, $V/2\pi=6$~MHz, and $\Delta_g/2\pi=4.8$~MHz. The decay profile exhibits a clear exponential form, $M_{\text{AFM}}^{\text{res}}(t) \propto e^{-\gamma t}$, over a significant duration. This behavior is a key signature of FVD, allowing us to extract the decay rate $\gamma$. 

The inset displays the extracted $\gamma$ (empty diamonds) alongside the results attained from numerical simulations for $N = 16$ without considering any experimental errors (gray line). We repeated the measurement for a larger $N=24$ ring (solid hexagons) and found that the decay rates collapse onto the same curve. This size-independence suggests a universal scaling behavior for a sufficiently long chain. Such behavior is consistent with the local nature of the decay dynamics: in the short-time regime, the Lieb-Robinson bound~\cite{1972_Lieb_FiniteGroupVelocity, 2012_Bloch_LightConeCorrelation, 2015_Eisert_OutOfEquilibrium} restricts the decay process to a local light-cone, making the overall decay rate insensitive to the total system size $N$. Our numerical simulations corroborate this, showing that $M_{\text{AFM}}^{\text{res}}(t)$ and $\gamma$ converge for $N > 12$ (except for very small $\Delta_l$).

Besides, Figure~\ref{fig_decay} shows that the FV decay rate diminishes as $\Delta_l$ decreases. This mechanism can be understood heuristically with the aid of Fig.~\ref{fig_setup}(b): the false vacuum's energy becomes resonant with product states containing a length-$L$ bubble when the bulk-energy gain $L\Delta_l$ compensates the surface-energy cost $V$. This defines a resonant bubble size $L_r \sim V/\Delta_l$. Since generating such a bubble requires a collective tunneling of $L_r$ spins, the nucleation rate $\gamma$ is expected to be exponentially suppressed with $L_r$. Indeed, instanton theory describing the FV decay of scalar quantum field~\cite{1977_Coleman_Instanton,2022_Devoto_FVDReview} and non-perturbative analytical study of the ferromagnetic Ising model~\cite{1999_Rutkevich_PerturbativeIsingModel} both predict an exponential suppression $\gamma\propto \exp(-\lambda V/\Delta_l)$, where $\lambda$ collects the remaining dependencies. For a standard Ising chain, $\lambda$ depends only on the transverse field $\Omega$~\cite{2021_Lagnese_FVDinSpinChains}. Intriguingly, our data in Fig.~\ref{fig_decay} exhibit a notable deviation from this exponential prediction, failing to collapse onto a straight line in the semi-log plot, especially at small $\Delta_l$.

\begin{figure}[!ht]
	\centering \includegraphics[width=0.94\columnwidth]{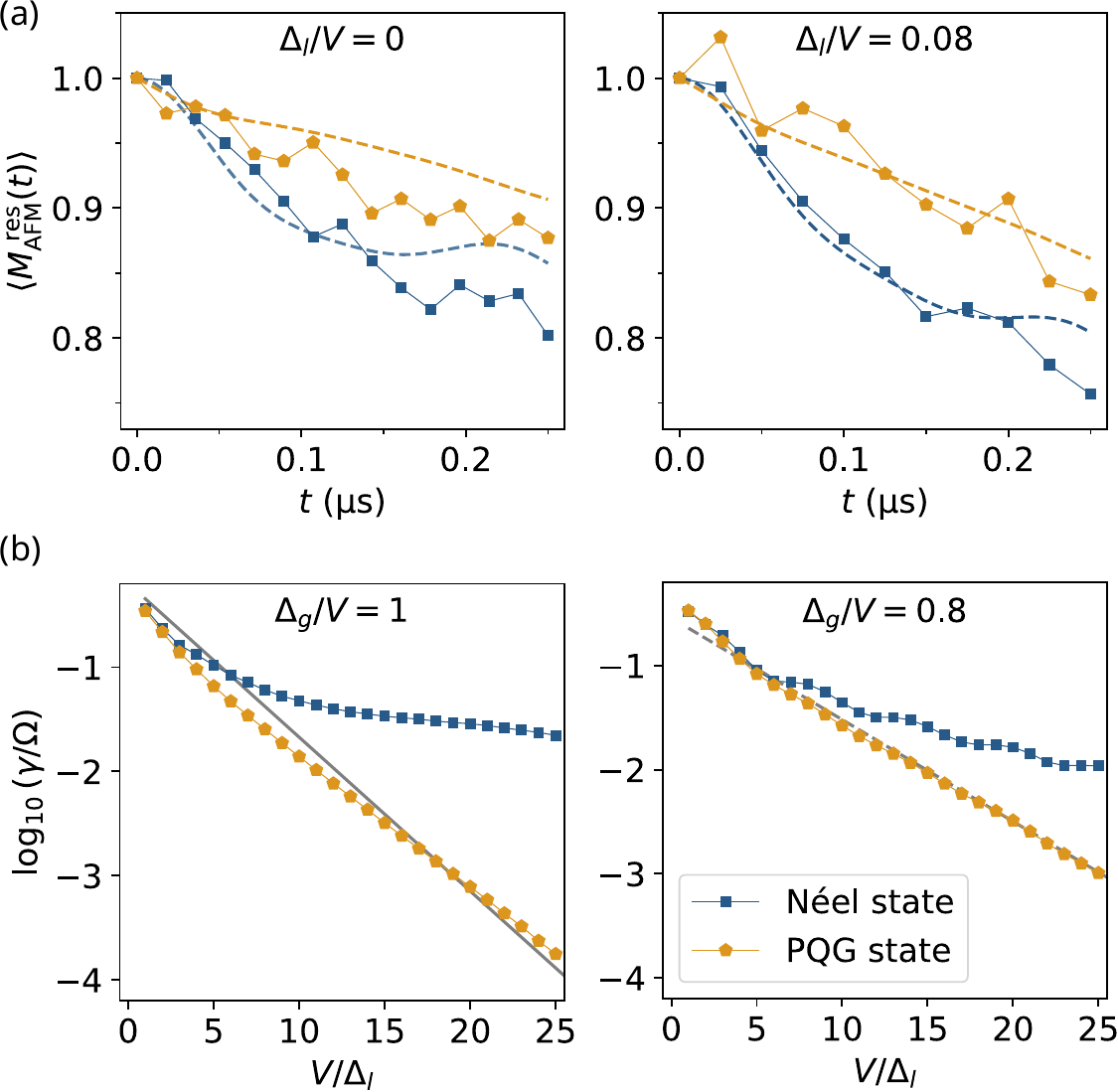}
	\caption{
    Comparison of FV decay of an initial Néel state (blue squares) versus the pre-quench ground (PQG) state (orange pentagons).
    (a) Measured evolution of the AFM order starting from the Néel state versus PQG state, for $\Delta_l/V=0$ (left) and $\Delta_l/V=0.08$ (right). Dashed lines are numerical simulations that include decoherence at the bench-marked level.
    (b) Comparison of decay rates, extracted from numerical simulation, for the Néel and PQG states as a function of $V/\Delta_l$. The left panel shows the result for $\Delta_g/V=1$ (nearest neighbor only), and the solid gray line is the theoretical prediction based on Ref.~\cite{2021_Lagnese_FVDinSpinChains}. The right panel shows the result for $\Delta_g/V=0.8$ (full van-der-Waals interactions), and the dashed gray line represents a linear fit for the PQG data.
    }
    \label{fig_NeelandFV}
\end{figure}

{\it Different initial states. ---}\label{sec.diff_init}
To understand the discrepancy observed in Fig.~\ref{fig_decay}, we investigate the decay dynamics from a different state: the pre-quench ground (PQG) state. It is the entangled ground state when $\Delta_l \to 0^-$ (i.e., in the opposite symmetry-broken sector) and therefore serves as a more faithful representation of the metastable false vacuum. Experimentally, we prepare the $|\rm PQG\rangle$ state by first initializing the system in the $|\text{Néel}\rangle$ state and then adiabatically ramping up $\Omega$ while holding $\Delta_l=0$ (see SM Sec.~\Romannum{1}~\cite{SM}). We then quench $\Delta_l$ to its target value and observe the decay process.

Figure~\ref{fig_NeelandFV}(a) directly compares the measured decay of the PQG state (orange pentagons) against the Néel state (blue squares) for both $\Delta_l/V = 0$ and $\Delta_l/V = 0.08$. The PQG state exhibits a significantly slower decay. This behavior is qualitatively confirmed by numerical simulations that incorporate decoherence (dashed lines in Fig. 3(a)) (see SM Sec.~\Romannum{2}~\cite{SM}). The different early-time dynamics of the two initial states can also be revealed by analytical analysis via a Baker-Campbell-Hausdorff (BCH) expansion (see SM Sec.~\Romannum{4}~\cite{SM}).

To gain deeper insight into these different dynamics, we numerically simulate the post-quench evolution of both states for various $\Delta_l$. We extract the decay rates $\gamma$ by fitting $M_{\text{AFM}}^{\text{res}}(t)$ over a $\Delta_l$-distinctive time window~\footnote{We extract $\gamma$ from numerical simulations by fitting $\ln [M_{\rm{AFM}}^{\rm{~res}}(t)]$ within a dynamically determined window $t \in [t_{\rm{start}}, t_{\rm{end}}]$, where $t_{\rm{start}} = 0.034+0.0058V/\Delta_l~(\rm{\mu s})$, $t_{\rm{end}} = 0.11+0.034V/\Delta_l~(\rm{\mu s)}$} exhibiting an exponential decay (see SM Sec.~\Romannum{3}~\cite{SM}). The results are shown in Fig.~\ref{fig_NeelandFV}(b) for two scenarios: (i) the standard Ising model ($\Delta_g = V$, nearest-neighbor interactions only),
for which an analytical expression for $\gamma$ is known for small longitudinal fields~\cite{1999_Rutkevich_PerturbativeIsingModel, 2021_Lagnese_FVDinSpinChains}, and (ii) the generalized Ising model ($\Delta_g = 0.8V$, $1/r^6$ van-der-Waals interactions beyond the nearest-neighbor), matching our experimental parameters from Fig.~\ref{fig_decay}. 

\begin{figure*}[ht!]
	\centering
    \includegraphics[width=1.85\columnwidth]{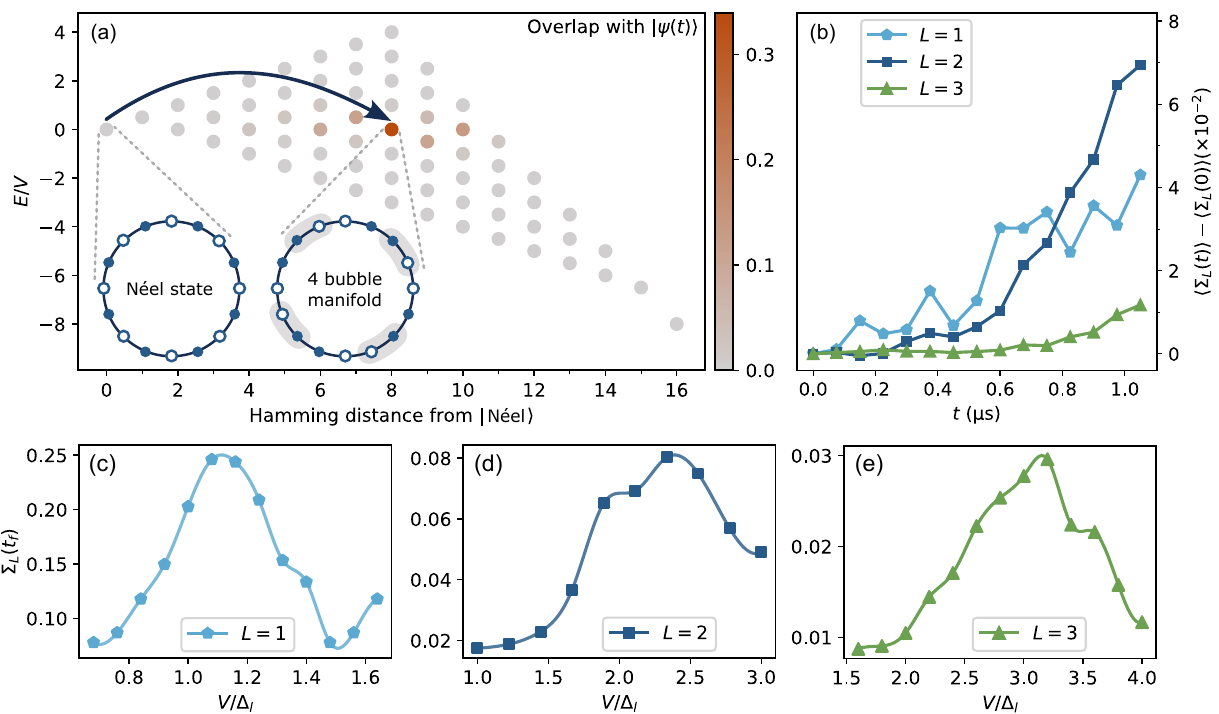}
	\caption{
     Resonant nucleation of true-vacuum bubbles. (a) Energy landscape of the product states at the $L=2$ resonance ($V/\Delta_l \approx 2$), on which the final state projection is drawn according to the color bar. Starting from the Néel state (lower left), the final state preferentially populating the $L=2$ bubble manifolds, with the most probable four-bubble manifold illustrated at the lower right.
     (b) Measured bubble density for sizes $L=1, 2, 3$ during the ramp (at $V/\Delta_l=2$), showing the steepest increase for the $L=2$ bubbles. (c-e) Experimental verification of the resonance condition. The final bubble densities $\langle\Sigma_L\rangle$ are plotted versus $V/\Delta_l$ for $L=1$ (c), $L=2$ (d), and $L=3$ (e), revealing distinct peaks at $V/\Delta_l \approx L$. Ramp parameters ($\Omega_f/(2\pi), T$) are independently optimized for each $L$ to maximize the signal: (c) ($1$~MHz, $0.5~\mu$s); (d) ($1.8$~MHz, $1~\mu$s); (e) ($1.8$~MHz, $2~\mu$s). 
    }
    \label{fig_resonance}
\end{figure*}

In the left panel of Fig.~\ref{fig_NeelandFV}(b), we compare the decay rate extracted from numerical simulations with theoretical predictions from Eq.~(6) of Ref.~\cite{2021_Lagnese_FVDinSpinChains} (solid gray line, without any fitting parameters). Remarkably, the decay rates for the PQG state show excellent agreement with the theory over four orders of magnitude, with no fitting parameters. In contrast, for the Néel state, it becomes difficult to identify a clear exponential decay regime when $V/\Delta_l \ge 10$. This is because the weak exponential decay signal is masked by stronger high-frequency oscillations (see SM Sec.~\Romannum{3}~\cite{SM}). Nevertheless, using the same time windows (as the PQG's) to fit $M_{\rm{AFM}}^{\rm{~res}}(t)$, we find that the resulted $\gamma$ for the Néel state rapidly deviate from the theoretical prediction as $\Delta_l$ decreases. These distinctions are striking given the high overlap between the two states, but can be understood heuristically: for small $\Delta_l$, the PQG state is, by construction, nearly an eigenstate of the post-quench Hamiltonian, ensuring its metastability; in contrast, the Néel state, which corresponds to the ground state when $\Delta_l \to +\infty$, represents a broad superposition of eigenstates of the post-quench Hamiltonian.

Scenario (ii) maps to a generalized Ising model with both global and staggered longitudinal fields, for which no analytical expression for $\gamma$ is known. Nevertheless, our simulations (Fig.~\ref{fig_NeelandFV}(b), right) show that the same dichotomy persists: the PQG state's decay rate still follows a clear exponential scaling, $\gamma \propto \exp(-\lambda V/\Delta_l)$ (dashed gray fit), albeit with a different slope $\lambda$, while the Néel state does not. This discovery suggests that the exponential suppression of $\gamma$ with symmetry-breaking field strength $\Delta_{l}$ is a universal feature of FVD in generalized Ising model, provided the system is prepared in a proper metastable state. Recent theoretical works have proved the rigorous theoretical bounds of this decay law in general short-range interacting systems~\cite{2025_YinChao_MetastableState}. However, the precise role of long-tail interactions in FVD dynamics, as present in our Rydberg platform, remains an open question.

{\it Resonant bubble nucleation. ---}
The early-time exponential-decay nucleation behavior we discuss so far is common to classical systems, continuous quantum fields, and discrete quantum systems. However, an isolated quantum system with a discrete energy spectrum, like our Ising chain, exhibits unique long-time dynamics. First, energy conservation in the unitary evolution prevents nucleated TV bubbles from expanding indefinitely to reach the global TV. Second, the discrete spectrum gives rise to resonant conditions that can facilitate bubble nucleation. This mechanism is akin to the resonant melting of prethermal states recently observed in a 2D Rydberg array~\cite{2025_Darbha_ResonantMelting}, driven by coupling to specific excitation manifolds.

We employ a different experimental protocol to observe the resonant nucleation. We first prepare the Néel state, and then ramp up $\Omega$ to a final value $\Omega_f$ over a duration $T \sim 1~\rm{\mu s}$ (see SM Sec.~\Romannum{1}~\cite{SM}). While this ramp is not strictly adiabatic, it is designed to transfer the state population into a more localized region in the Hilbert space, which helps reveal the underlying nucleation resonances more clearly. Figure~\ref{fig_resonance} illustrates the system's response to this ramp protocol. Figure 4(a) depicts the energy landscape of the product states at $V/\Delta_l \approx 2$, the resonance condition for $L=2$ bubbles. Each dot in the landscape represents a manifold of product states, grouped by their Hamming distance from the $|\text{Néel}\rangle$ state and their relative static energy $E/V$ (at $\Omega=0$). The color map shows the simulated projection probability of the final state, which evolves from $|\text{Néel}\rangle$ after a $T=1~\mu\text{s}$ ramp to $\Omega_f/2\pi=1~\text{MHz}$. The evolution, constrained by energy conservation, preferentially populates the $L=2$ bubble manifolds (located on the zero-potential-energy line). This constraint implies that for longer evolution, the dynamics is dominated by the nucleation and annihilation of $L=2$ bubbles.

To quantify the bubble nucleation, we measure the bubble density observables $\hat{\Sigma}_L$ for TV bubbles of length $L$,
\begin{subequations}\label{eq.true_bubble_density}
\begin{align}
    \hat{\Sigma}_1 & = \frac{1}{N}\left( \sum_{i\in \text{odd}} \hat{n}_{i}\hat{n}_{i+1}\hat{n}_{i+2} + \sum_{i\in \text{even}} \hat{g}_{i}\hat{g}_{i+1}\hat{g}_{i+2} \right), \nonumber\\
    \hat{\Sigma}_2 & = \frac{1}{N}\left( \sum_{i\in \text{odd}} \hat{n}_{i}\hat{n}_{i+1}\hat{g}_{i+2}\hat{g}_{i+3} + \sum_{i\in \text{even}} \hat{g}_{i}\hat{g}_{i+1}\hat{n}_{i+2}\hat{n}_{i+3}\right),\hskip -12pt\nonumber\\
    \hat{\Sigma}_3 & = \frac{1}{N}\left( \sum_{i\in \text{odd}} \hat{n}_{i}\hat{n}_{i+1}\hat{g}_{i+2}\hat{n}_{i+3}\hat{n}_{i+4} + \sum_{i\in \text{even}} \hat{g}_{i}\hat{g}_{i+1}\hat{n}_{i+2}\hat{g}_{i+3}\hat{g}_{i+4} \right),\nonumber
\end{align}
\end{subequations}
with site index $i \in [1,N]$ and periodic boundary conditions for the ring. As an example, Fig.~\ref{fig_resonance}(b) plots the measured time evolution of $\langle\hat{\Sigma}_L(t)\rangle$ during a ramp to $\Omega_f / 2\pi = 1.8~\text{MHz}$ under the $L=2$ resonance condition ($V / 2\pi=\Delta_g / 2\pi = 10$~MHz, $\Delta_l / 2\pi = 5$~MHz). As expected, the $L=2$ bubble density shows the most significant growth, clearly outpacing the growth of $L=1$ and $L=3$ bubbles. By changing $\Omega_f$ and $T$, we can selectively enhance bubbles of different lengths. Figures ~\ref{fig_resonance}(c-e) plot the measured bubble densities $\langle\hat{\Sigma}_L\rangle$ after the $\Omega$-ramp. As expected, we observe pronounced resonance peaks for $L=1$, $2$, and $3$ centered near $V/\Delta_l \approx 1$, $2$, and $3$, respectively.

{\it Summary and outlook. ---}
We simulate the false vacuum decay and bubble nucleation dynamics in a generalized 1D AFM Ising ring using a Rydberg atom array. A key finding is the strong initial-state dependence of the FVD: a proper metastable (PQG) state exhibits a cleaner and more extended exponential decay of the AFM order, with a rate that follows the expected exponential suppression with the inverse staggered field, whereas a simple Néel state does not. We also probe the resonant bubble nucleation process unique to discrete systems. In future, one may explore non-local observables and further clarify the role of long-range interactions in FVD dynamics. Our work paves the way for future studies of many-body tunneling on Rydberg platforms, including systems with multiple true and false vacuums that can be realized in $\mathbb{Z}_3$-symmetry-broken phases~\cite{Bernien_2017,2022_Takacs_tricritical}, as well as models of complex lattice geometries~\cite{2021_Semeghini_SpinLiquid,2025_Lukas_PrethermalGauge}.

{\it Note added. ---} During the preparation of this manuscript, we became aware of a related effort in a 2D Rydberg array by the group of Prof. Christian Gross at the University of Tübingen~\cite{2025_Gross_CollectiveNucleation}.

\begin{acknowledgments}
{\it Acknowledgements. ---} This work is supported by the National Natural Science Foundation of China (NSFC) (Grants No. 92565306, No. 92265205, No. 12234012, and No. W2431002), and the Quantum Science and Technology--National Science and Technology Major Project (2021ZD0302104). R.L. is supported by the ``Gravitational Wave Detection'' program (2023YFC2205800) funded by the Ministry of Science and Technology of the People's Republic of China. X.W. at Cornell University is supported by the U.S. Department of Energy through Award Number: DE-SC0023905.
\end{acknowledgments}

\bibliographystyle{apsrev4-2}
%

\clearpage
\onecolumngrid

\setcounter{secnumdepth}{2}  
\pagenumbering{arabic}   
\setcounter{page}{1}

\renewcommand{\theequation}{S\arabic{equation}}
\setcounter{equation}{0}

\renewcommand{\thefigure}{S\arabic{figure}}
\setcounter{figure}{0}

\renewcommand{\thesection}{\Roman{section}}
\renewcommand{\thesubsection}{\Roman{section}.\Alph{subsection}}
\setcounter{section}{0}

\makeatletter
\renewcommand\section{\@startsection{section}{1}{0pt}%
  {24pt plus 6pt minus 6pt}
  {16pt plus 6pt minus 4pt}
  {\centering\normalfont\normalsize\bfseries}} 
\makeatother

\begingroup
\centering
\large\bfseries Supplemental Material: \\
Probing false vacuum decay and bubble nucleation in Rydberg atom array\par
\vspace{6pt}
\normalfont\normalsize   
Yu-Xin Chao$^{1,2,*}$, Peiyun Ge$^{1,2,*}$, Zhen-Xing Hua$^{1}$, Chen Jia$^{1}$, Xiao Wang$^{3}$, Xinhui Liang$^{1}$, Zongpei Yue$^{1}$, Rong Lu$^{1}$,\\ Meng Khoon Tey$^{1,4,5}$, Xiao Wang$^{1,2,6}$, and Li You$^{1,2,4,5}$\par
\vspace{2pt}
\textit{$^{1}$State Key Laboratory of Low Dimensional Quantum Physics,\\ Department of Physics, Tsinghua University, Beijing 100084, China}\\
\textit{$^{2}$Beijing Academy of Quantum Information Sciences, Beijing 100193, China}\\
\textit{$^{3}$Department of Physics, Cornell University, Ithaca, NY, USA}\\
\textit{$^{4}$Frontier Science Center for Quantum Information, Beijing 100084, China}\\
\textit{$^{5}$Hefei National Laboratory, Hefei, Anhui 230088, China}\\
\textit{$^{6}$Department of Physics, University of Oxford, Oxfordshire, UK}\\\par
\vspace{8pt}
\justifying
In this Supplemental Material, we provide technical details supporting the experimental and theoretical results presented in the main text.
Section~\ref{SM.sec.ExperimentalProcedures} details the experimental implementation of the staggered longitudinal field using spatial light modulators and outlines the time sequences for state preparation and measurement.
In Section~\ref{SM.sec.ExperimentalImperfections}, we characterize experimental imperfections, including state preparation and measurement (SPAM) errors and decoherence, and describe their incorporation into our numerical modeling.
Section~\ref{SM.sec.Decay_vs_deltal} elaborates on the analysis of exponential decay dynamics, verifying the robustness of the extracted decay rates with two different fitting protocols.
Finally, Section~\ref{SM.sec.BCH} presents a Baker-Campbell-Hausdorff (BCH) expansion of the antiferromagnetic order parameter, providing an analytical derivation of the early-time dynamics that distinguishes the stability of the pre-quench ground (PQG) state from the Néel state.

\par
\endgroup
\vspace{12pt}

\section{Experimental procedures}\label{SM.sec.ExperimentalProcedures}

\subsection{Realization of staggered longitudinal field}
In our setup, a spatial-light-modulator (SLM) shapes an 830-nm laser beam to create an optical tweezer array for trapping the atoms. This trapping light is turned off abruptly prior to the main experimental sequence. A staggered longitudinal field, $\Delta_{\rm{stagger}}=(-1)^j\Delta_l$, is produced using a second SLM, which creates an addressing array of 1014-nm tweezers targeting even-numbered atoms (j=even) in the ring. Due to the large detuning from the Rydberg transition ($|6P_{3/2}\rangle \rightarrow |70S_{1/2}\rangle$), the resulting AC Stark shifts of addressing tweezers, $\Delta_{\text{add}} = \Delta_{\uparrow} - \Delta_{\downarrow}$, is dominated by the ground-state component $\Delta_{\downarrow}$. For a beam waist of $1.09~\mu\text{m}$, an addressing power of $\sim 15.5$~mW yields a shift of $\Delta_{\text{add}} = 30$~MHz. To achieve the targeted overall detuning of $-\Delta_{g}+(-1)^{j}\Delta_{l}$ for the j-th atom, we set the two-photon detuning of the 1013-nm and 420-nm lasers, which are applied globally to the atoms, to $\Delta=\omega_{\rm1013nm}+\omega_{\rm420nm}-\omega_{\uparrow\downarrow}=-\Delta_{g}-\Delta_{l}$ and set the addressing shift to $\Delta_{\rm  add}=2\Delta_{l}$.

To ensure that all addressed atoms experience the same AC Start shift, we need to ensure good overlap between the 1014-nm addressing tweezers and the 830-nm trapping tweezers, and also achieve the same intensity for each addressing tweezer. 

For the first step, alignment, we adapt the protocol from Ref~\cite{thesis_Gabriel} with minor modifications. We begin by cooling the atoms near their vibrational ground state, ensuring they are localized at the center of the 830-nm tweezers. We then apply an additional blazed grating to the 1014-nm SLM's initial phase pattern to shift the addressing tweezer positions, subsequently measuring the response of atoms. This response allows us to infer the relative position between the atoms and the 1014-nm tweezers. Our modification to the original protocol is to measure atom loss probability—by abruptly turning the addressing tweezers on and off—instead of measuring the AC Stark shift of the addressing tweezers. This loss probability exhibits a specific dependence on alignment: it is near zero for perfect overlap, increases for slight misalignment, and returns to zero when completely non-overlapped. By sweeping the blazed grating and monitoring this loss signal, we can determine the optimal addressing tweezer position. This information is fed back to generate a corrected 1014-nm SLM phase pattern. Several rounds of this feedback reduce the overall loss probability to below 3\%, indicating good overlap between the two tweezer sets.

The second step addresses intensity uniformity via feedback. We measure the AC Stark shift induced by each addressed tweezer and use this information to modify the corresponding tweezer's weight in the SLM phase pattern. The final non-uniformity of the AC Stark shifts across the addressing tweezers is approximately 2.8\% (rms).

\subsection{Experimental sequence}
\begin{figure}[!ht]
	\centering \includegraphics[width=0.95\columnwidth]{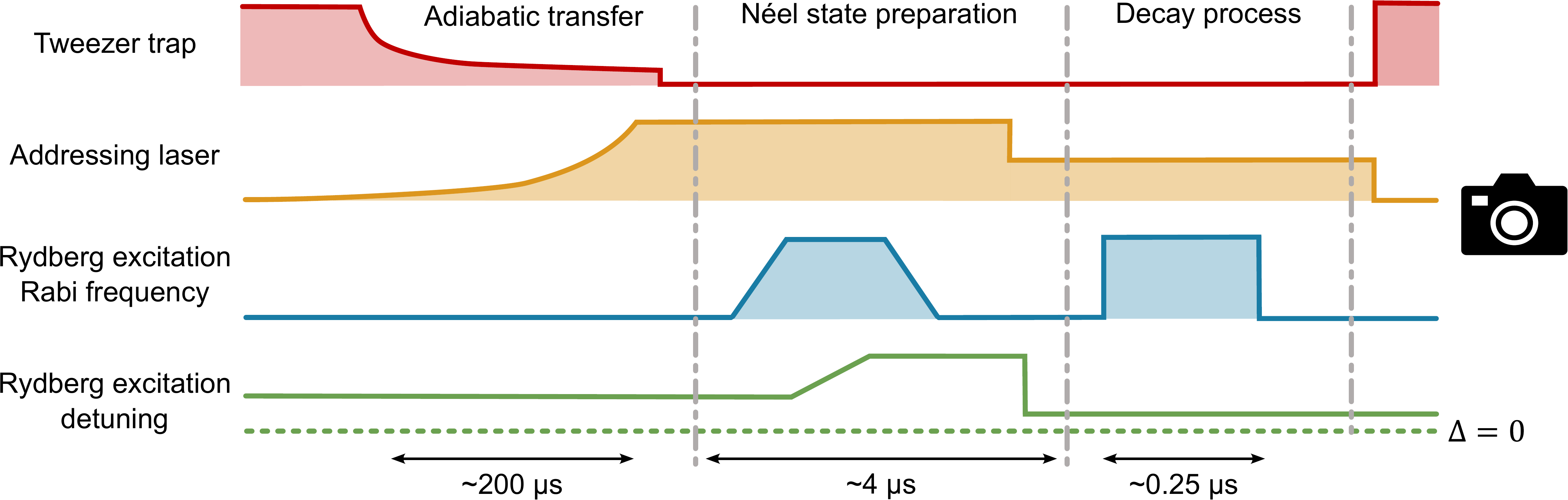}
	\caption{Experimental time sequence for Néel state preparation and subsequent decay measurement. The sequence consists of adiabatic transfer, Néel state preparation via a Landau-Zener sweep, the decay process under the target Hamiltonian, and the final projective measurement.
    }
    \label{fig_TimeSequence_Neel}
\end{figure}

Prior to the main sequence, atoms are arranged in a ring configuration with interatomic distances of $6.65$ or $7.24~\mu\text{m}$, corresponding to nearest-neighbor interactions of $10$ or $6~\text{MHz}$, respectively. The atoms are cooled to $\sim 3~\mu\text{K}$ and initialized in the ground state $|5S_{1/2}, F=2, m_F=2\rangle$. A $30~\text{G}$ magnetic field is applied parallel to the propagation direction of the Rydberg excitation beams. Whenever there is a residual offset between the center of any 1014-nm addressing tweezer and that of the 830-nm trapping tweezer, the atom, which is confined originally in the trapping tweezer, would experience a larger fluctuation in $\Delta_{\rm add}$ and also stronger heating if we abruptly turn on the addressing tweezer. To avoid this problem, we perform an ``adiabatic transfer'' just before the core experiment. This involves simultaneously ramping down the 830-nm tweezers while ramping up the 1014-nm addressing tweezers  at the even-site until $\Delta_{\rm add}/2\pi=30$~MHz  (see Fig.~\ref{fig_TimeSequence_Neel}). At the end, atoms in the even sites are adiabatically transferred into the 1014-nm addressing tweezers while those in the odd sites remain at the centers of the trapping lasers.

\begin{figure}[!ht]
	\centering \includegraphics[width=0.7\columnwidth]{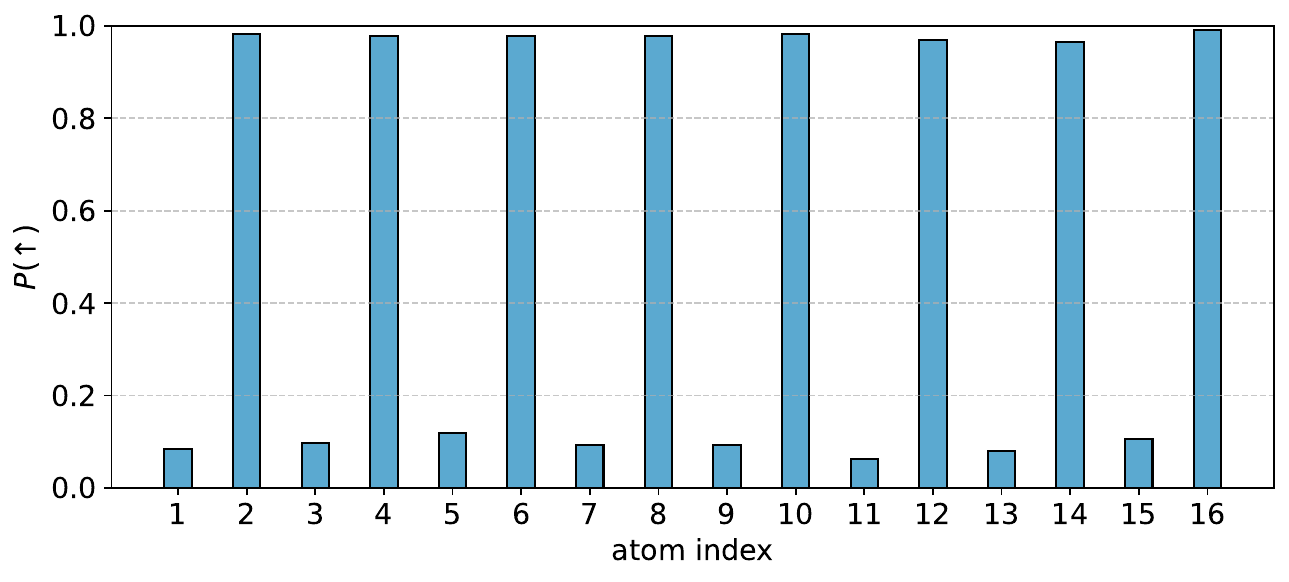}
	\caption{Measured site-resolved Rydberg probability $P(\uparrow)$ for a 16-atom Néel state, without detection error correction.
    }
    \label{fig_Neel_Preparation}
\end{figure}

To prepare the $|\text{Néel}\rangle$ state, we then abruptly turn off the addressing laser and perform a Landau-Zener sweep by tuning the Rydberg laser detuning $\Delta/2\pi$ from 25.7~MHz to 33.1~MHz, with a Rabi frequency $\Omega/2\pi \sim 1$~MHz. This process excites all even-site atoms to the Rydberg state, leaving odd-site atoms in the ground state. Figure.~\ref{fig_Neel_Preparation} shows the measured probability for each atom to be excited to the Rydberg state after the Landau-Zener sweep. To measure the subsequent decay dynamics of $|\text{Néel}\rangle$ state, we set the addressing laser intensity and Rydberg laser detuning to their target values, and then turn on the Rabi frequency $\Omega/2\pi=1.8~\rm{MHz}$, allow the system to evolve for a variable time, and finally turn off the Rydberg lasers and measure the state of each atom. The details of the sequence are illustrated in Fig.~\ref{fig_TimeSequence_Neel}. 

\begin{figure}[!ht]
	\centering \includegraphics[width=0.95\columnwidth]{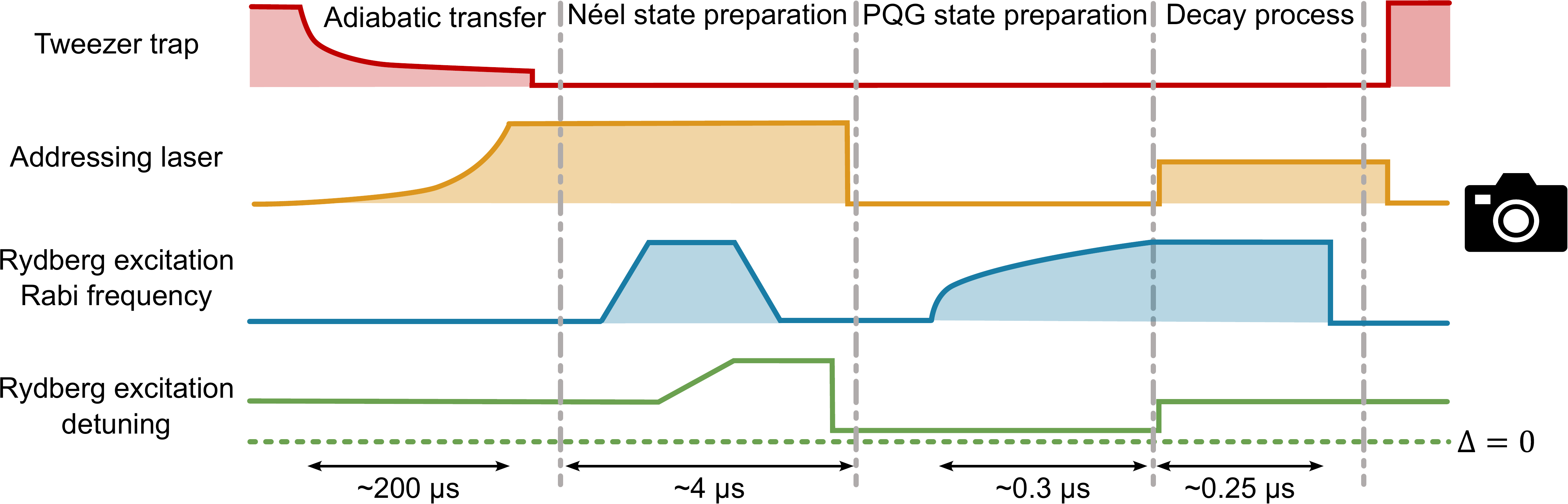}
	\caption{Experimental time sequence for PQG state preparation and decay measurement. This sequence has an extra ``PQG state preparation'' stage in comparison to Fig.~\ref{fig_TimeSequence_Neel}.
    }
    \label{fig_TimeSequence_FV}
\end{figure}

As discussed in the main text (Fig.~3(a)), we contrast the decay dynamics of the Néel state with the entangled pre-quenched ground (PQG) state. The corresponding time sequence for the PQG experiment is shown in Fig.~\ref{fig_TimeSequence_FV}. Compared with the  Néel-state sequence in Fig.~\ref{fig_TimeSequence_Neel}, it has an extra "PQG state preparation" stage. Following the Néel state preparation, we set the parameters to $\Delta_{g}/2\pi=\Delta_{g,\rm{quench}}/2\pi=4.8~\rm{MHz}$ and $\Delta_{l}=0$. This is achieved by turning off the addressing laser ($\Delta_{\rm{add}}=0$) and tuning the two-photon detuning to $\Delta=-\Delta_{g,\rm{quench}}$. Afterwards, we ramp up the Rydberg laser Rabi frequency $\Omega/2\pi$ from 0 to $1.8~\rm{MHz}$ within $0.3~\mu$s using a lineshape $\propto \sqrt{t}$ ($t$ being the time).

The experiment described in Fig.~4 of the main text, which detects resonant bubble nucleation, also begins with the Néel state. The sequence mirrors Fig.~\ref{fig_TimeSequence_Neel}, differing only in how the decay process is initialized. Here, $\Omega$ is ramped using a lineshape $\propto \sqrt{t}$ from 0 to $\Omega_{f}/2\pi \sim 1~\rm{MHz}$, instead of being applied abruptly.

\section{Experimental imperfections}\label{SM.sec.ExperimentalImperfections}
Discrepancies between experimental data and raw numerical simulation results arise from several experimental imperfections. In this section, we report the calibration of these error sources and describe how we include or compensate for them in our numerical modeling.

\subsection{SPAM error}
We estimate the state preparation fidelity for the ground state $|\downarrow\rangle=|5S_{1/2}, F=2, m_{F}=2\rangle$ by measuring the microwave $\pi$-pulse fidelity between $|\downarrow\rangle$ and $|5S_{1/2}, F=1, m_{F}=1\rangle$. This fidelity is $>98.5\%$, providing a lower bound for the $|\downarrow\rangle$ preparation fidelity.

For state detection, we perform single-atom-resolved measurement by switching the 830-nm trapping tweezers back on. Ground state atoms are recaptured and subsequently detected via fluorescence imaging, while Rydberg atoms are lost. To enhance detection fidelity, we apply a 3~µs, 3~W, 8.2~GHz detection-facilitation microwave pulse during the recapture process. This pulse transfers atoms from the $|70S\rangle$ state to neighboring Rydberg states with longer lifetimes. Through single-atom calibration, we quantify the measurement errors as follows: the probability of misidentifying a Rydberg state as a ground state ($|\uparrow\rangle \rightarrow |\downarrow\rangle$) is $p_{1}\sim4\%$, and the probability of misidentifying a ground state as a Rydberg state ($|\downarrow\rangle \rightarrow |\uparrow\rangle$) is $p_{2}\sim1.5\%$. Notably, the sole effect of these SPAM errors on the AFM order parameter is an overall multiplicative factor $(1-p_{1}-p_{2})$, which is normalized out in the rescaled AFM order.

\subsection{Decoherence}
The numerical simulations in Fig.~3(a) incorporate single-atom decoherence, which includes two main sources: (1) Depolarization: modeled by a longitudinal relaxation time $T_{1}=28\rm{\mu s}$. This value, obtained from single-atom fitting, accounts for the Rydberg state lifetime, off-resonant scattering from the intermediate $|6P_{3/2}\rangle$ state, and the leakage of detection-facilitation microwave. (2) Dephasing: characterized by measured values of $T_{2}^{*}=3.8 \rm{\mu s}$ and $T_{2}=11 \rm{\mu s}$.

To model the experimental system with these errors, we make use of the Lindblad master equation~\cite{BreuerPetruccione,Manzano2020}. For each site $j$ we include a decay jump operator
\begin{equation}
    L_{\rm{depolarization},j}=\sqrt{\gamma_1}\sigma^-_j,
\end{equation}
and a dephasing jump operator
\begin{equation}
    L_{\rm{dephasing},j}=\sqrt{\gamma_2}\sigma_j^z,
\end{equation}
with $\gamma_1=\frac{1}{T_1}$ and $\gamma_2=\frac{1}{2T_2^*}$.

For the Néel state simulation, we initialize the system in the ideal Néel product state and evolve it with this master equation. For the PQG state simulation, we also begin with the Néel state, first simulating the PQG state-preparation sequence with dissipation to obtain the initial state, and then simulating the subsequent quench dynamics using the same master equation.

\section{The dependence of FVD behavior on symmetry-breaking field}\label{SM.sec.Decay_vs_deltal}

\subsection{Theoretical predictions of semiclassical and quantum spin model}

The decay rate of the false vacuum depends on the staggered field strength $\Delta_{l}$. Below, we quantitatively analyze how the decay of the rescaled AFM order depends on $\Delta_l$. We focus on deviations between our numerical simulation results and the theoretical predictions.

From the semiclassical perspective, false-vacuum decay is governed by an instanton (bounce) process \cite{1969_Langer_MetastableState,1977_Coleman_Instanton,1977_Callan_1stQuantumCorrection}: a bubble of the true vacuum nucleates within the false vacuum, and the decay rate is controlled by the Euclidean action of this saddle-point trajectory. In the thin-wall limit, the action scales linearly with the resonant bubble size $L_r\sim V/\Delta_l$, leading to a decay rate of the form $\Gamma \sim \frac{\Delta_l}{V}e^{-\lambda V/\Delta_l}$, with prefactor determined by fluctuations around the instanton.

References~\cite{1999_Rutkevich_PerturbativeIsingModel} give analytical derivations for the FV decay rate in a ferromagnetic Ising model with both transversal and global longitudinal fields. This model directly maps to our anti-ferromagnetic Ising model with both transversal and staggered longitudinal fields when $\Delta_g=V$. They show that in the thin-wall limit (where the resonant bubble size $V/\Delta_l \gg 1$), the exponential decay rate of return probability (associated with creating a single bubble of resonant size) obeys
\begin{equation}
    \Gamma\propto\frac{\Delta_l}{V}\exp(-\lambda\frac{V}{\Delta_l}).
    \label{eq.decayrate}
\end{equation}
The decay rate of AFM order should be considered as decay rate of return probability $\Gamma$ multiplied by the resonant bubble size $L_r$,
\begin{equation}
    \gamma\propto L_r\frac{\Delta_l}{V}\exp(-\lambda\frac{V}{\Delta_l})=\exp(-\lambda\frac{V}{\Delta_l}).
    \label{eq.decayrate}
\end{equation}
Although initially derived in the thin-wall limit, this result's validity can be extended up to $V/\Delta_l\gtrsim1.5$ by including further instanton corrections~\cite{2000_Munster_BeyondThinWall}. Accounting for lattice effects further refines the exponent and the prefactor predicted by the continuum theory~\cite{Lagnese2024}.

\subsection{Extraction of FV decay rate}

\begin{figure}[!ht]
	\centering	\includegraphics[width=0.95\columnwidth]{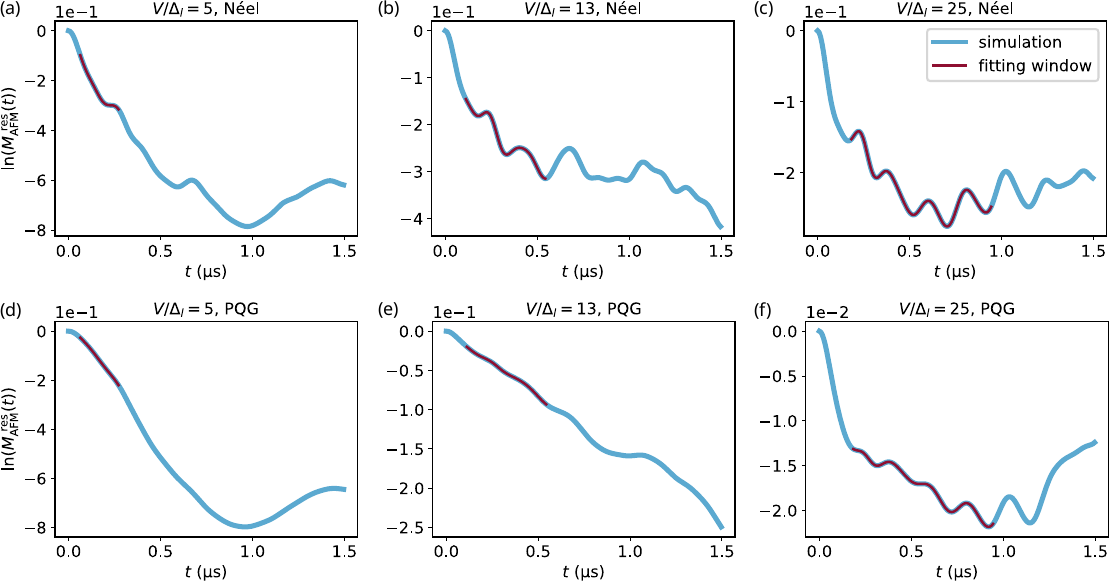}
	\caption{Simulated decay of rescaled AFM order in an $N=16$ ring with parameters $\Omega/2\pi=1.8$~MHz, $V/2\pi=6$~MHz, and $\Delta_g/2\pi=4.8$~MHz with different $V/\Delta_l$. The light blue lines are the numerical simulation results, and the dark red lines mark the $M_{\rm{AFM}}^{\rm{~res}}(t)\propto e^{-\gamma t}$ fitting time window.
    }
    \label{fig_SM_Fitting}
\end{figure}

\begin{figure}[!ht]
	\centering	\includegraphics[width=1\columnwidth]{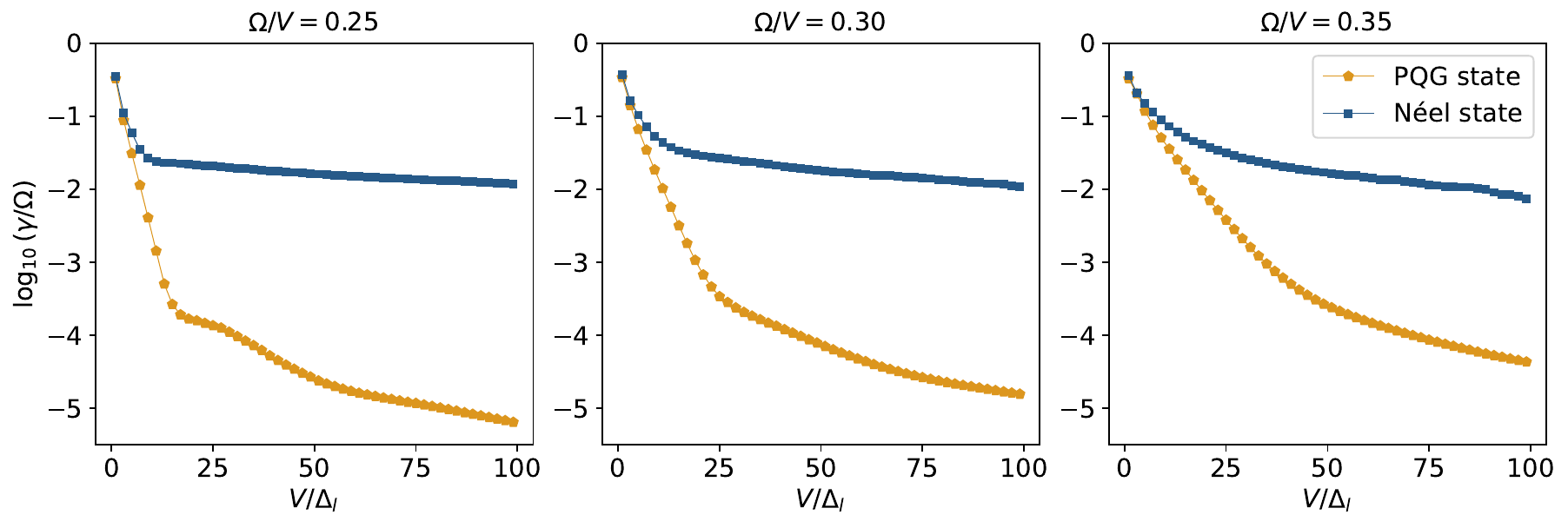}
	\caption{Numerical simulations for an infinite Rydberg chain with $V/2\pi=6~\mathrm{MHz}$ and $\Delta_g/2\pi=6~\mathrm{MHz}$. We extract the AFM-order decay rate $\gamma$ (normalized by $\Omega$) and fit its exponential scaling as a function of $V/\Delta_l$ for different Rabi frequencies $\Omega$. Using a fixed observation time $T=4\mu s$ and the fitting procedure used in the main text, the PQG initial state shows a wider linear regime in $\log_{10}(\gamma/\Omega)$ versus $V/\Delta_l$ as $\Omega$ increases. In contrast, results of the N\'eel initial state depends relatively weakly on $\Omega$.
    }
    \label{fig_mps_infinite}
\end{figure}

\begin{figure}[!ht]
	\centering	\includegraphics[width=1\columnwidth]{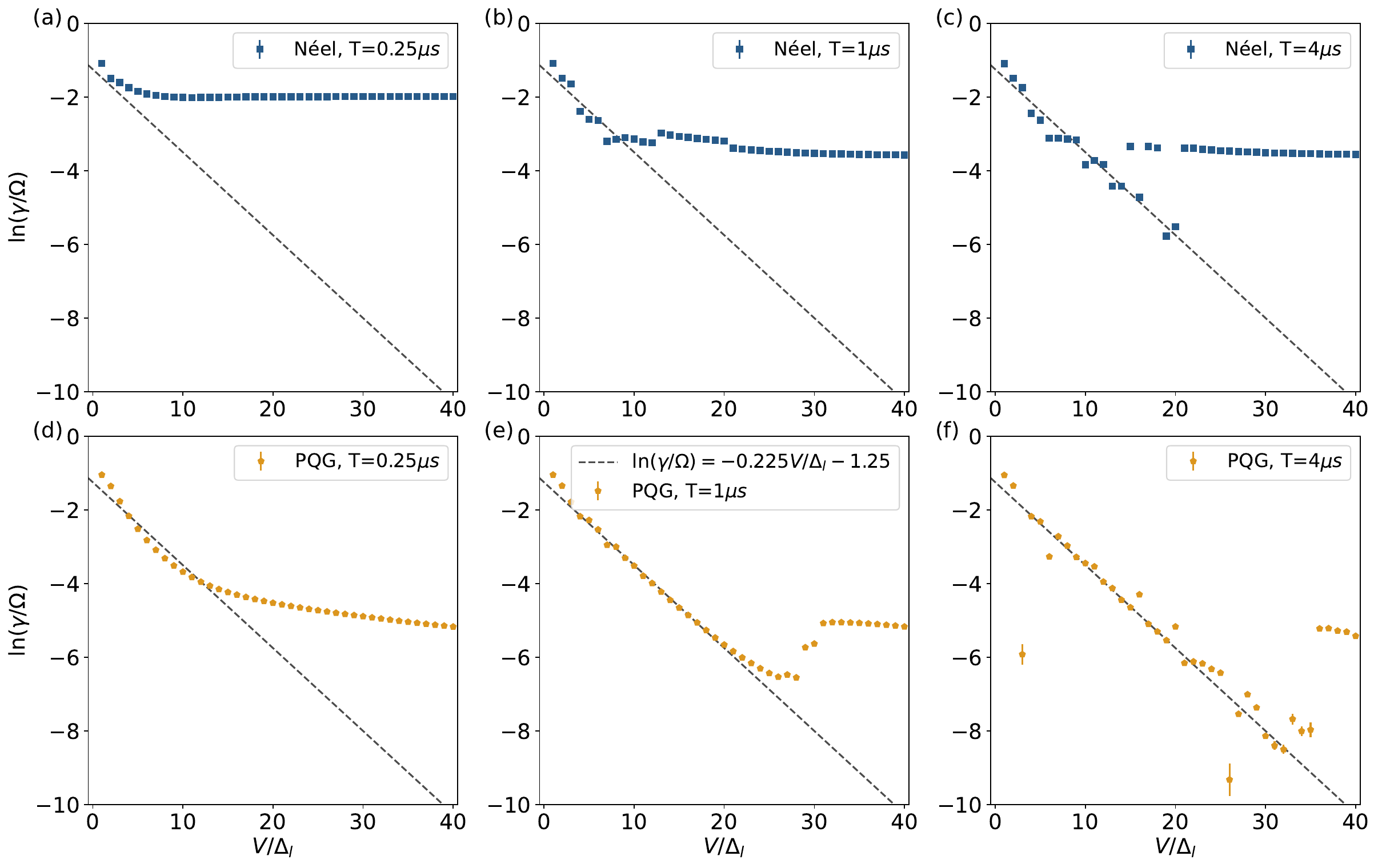}
	\caption{Numerical simulations for $N=16$ Rydberg ring at parameters $\Omega/2\pi=1.8$~MHz, $V/2\pi=6$~MHz, and $\Delta_g/2\pi=4.8$~MHz. Fitted AFM-order decay rate $\gamma$ under different $T$ (see text), starting from the two initial states $|\text{Néel}\rangle$ and $|\text{PQG}\rangle$. The extracted decay rates $\gamma$ generally shows an exponential scaling with $V/\Delta_l$. The dashed lines in all subplots are identical, and are the fit to the results in (f). 
    \textbf{Finite fitting-time effect:}
    As $V/\Delta_l$ increases, the decay timescale $\sim e^{V/\Delta_l}$ grows exponentially; thus, fits within a fixed window $T$ fail to track the dashed-line trend for small $\Delta_l$. From (a) $\to$ (c) [and (d) $\to$ (f)], extending the observation window $T$ postpones the onset of deviations to larger $V/\Delta_l$.
    }
    \label{fig_smalldl}
\end{figure}

Figure~\ref{fig_SM_Fitting} shows several examples of time evolution of AFM order, from different initial states, at different staggered fields $\Delta_l$. Focus first on the N\'eel state [sub-figures (a)-(c)]. In the initial stage, the AFM order decays at a similar rate for different values of $\Delta_l$, with $M_{\rm{AFM}}^{\rm{~res}}(t)$ decaying from 1 to about $\rm{e}^{-0.1}$. This period is known as the quantum Zeno regime~\cite{2008_Facchi_QuantumZeno}. After the initial period, it enters into the FVD regime, which we are interested in. For N\'eel state however, the FVD dynamics is accompanied by persistent oscillations. This leads to difficulty in pin-pointing the FVD decay regime and in extracting the decay rate, especially when $\Delta_l$ is small.

In contrast, the PQG state [sub-figures (d)-(f)] has a very different behavior. In the quantum Zeno regime, the initial drop becomes smaller as $\Delta_l$ is reduced. For $V/\Delta_l \lesssim 20$ the quantum Zeno regime and the FVD regimes connect smoothly with the same gradients [sub-figures (d) and (e)]. For smaller $\Delta_l$, the AFM order shows a distinctive multi-step decay [sub-figure (f)], with FVD regime lying between the quantum Zeno regime and the regime showing persistent oscillations. It should be noted that, compared to the N\'eel state, the FVD regime of the PQG state shows a much cleaner exponential decay over an extended period. This is a specific signature similar to the FV decay in the quantum field.

To extract the decay rates shown in Fig.~3(b) in the main text, we use a $\Delta_l$-dependent experienced-based fitting time window $t \in [t_{\rm{start}}, t_{\rm{end}}]$, where $t_{\rm{start}} = 0.034+0.0058V/\Delta_l~(\rm{\mu s})$, $t_{\rm{end}} = 0.11+0.034V/\Delta_l~(\rm{\mu s)}$. As shown in sub-figures (d-f), the resulting fitting windows (marked by the dark red lines) roughly capture the exponential FVD regime for the PQG state. That is, we only use the simulated data points within this window to fit $M_{\rm{AFM}}^{\rm{~res}}(t)\propto e^{-\gamma t}$. For the N\'eel state, we use the same time window as PQG's for extracting the decay rate, since it is difficult to distinguish the FVD regime. 

In Fig.~3(b) of the main text, we show that the fitted decay rate $\gamma$ of PQG state follows an excellent agreement with the theoretical prediction for a 1D ferromagnetic Ising model [Eq. (6) of ref.~\cite{2021_Lagnese_FVDinSpinChains}] for $V/\Delta_l \in [1,25]$. In fact, when $\Delta_l$ is very close to 0, the extracted $\gamma$ of the PQG state using the aforementioned fitting method deviates from that prediction. This is because the exceptionally slow and small FV decay at small $\Delta_l$ makes it challenging to isolate an exponential fitting window free from the post-quench oscillations~\cite{Lagnese2024}. As a result, even though the intrinsic FVD decay rate is exponentially suppressed with $V/\Delta_l$, these oscillations effectively inflate the fitted $\gamma$.

The mixing of FVD signal and post-quench oscillations can be mitigated by using a larger transverse field $\Omega$. A larger $\Omega$ (provided it remains sufficiently far from the critical point $2\Omega/V=1$) accelerates the FVD process, thereby allows for better extraction of the FVD signal~\cite{2021_Lagnese_FVDinSpinChains}. To show the influence of $\Omega$, we simulate the FVD of an infinite Rydberg chain with nearest-neighbor interactions for different Rabi frequencies $\Omega$ using infinite-MPS time evolution~\cite{Vidal2007iTEBD,tenpy} with bond dimension $D=100$. The results are shown in Fig.~\ref{fig_mps_infinite}. For the PQG state, the  $\ln\gamma$ varies linearly with respect to $V/\Delta_l$ for $V/\Delta_l\in[1,21]$ for $\Omega/V=0.3$ (corresponding to our experimental parameter). Increasing the Rabi frequency to $\Omega/V=0.35$ extends the linear regime to about $[1,35]$, whereas decreasing it to $\Omega/V=0.25$ shrinks the linear regime to about $[1,13]$. For the N\'eel state, all three values of $\Omega$ yield nearly the same decay behavior.

In the end, we comment that for small $\Delta_l$ in a finite system, the resonant bubble size may become comparable to or greater than the system's length. Under that condition, a bubble-size blockade effect effectively forbids the nucleation~\cite{ge2025twostate}.

\subsubsection{Verification of the results by a different fitting method}
As discussed in the section above, the decay rates of Fig.~3(b) in the main text are obtained using a $V/\Delta_l$-dependent experienced-based fitting time window. To verify the robustness of our conclusions, we also examine an alternative fitting method based on a decay-percentage criterion, yielding qualitatively similar conclusion. The fitting window is chosen as follows. First, we find the minimum of the function $y(t)=M^{~\rm{res}}_{\rm{AFM}}(t)$ within a fixed time interval $t\in[0,T]$ ($y_{\text max}=y(0)=1$ by definition). We then select a single continuous section of data $y(t)\in[y_{\text{low}},y_{\text{high}}]$ where $y_{\text{low(high)}}=y_{\text{min}}+\alpha_{\text low(high)}(y_{\text{max}}-y_{\text{min}})$.

The fitting results are shown in Fig.~\ref{fig_smalldl} for the two initial states and different $T$, setting $\alpha_{\text low}$= 0.2 and $\alpha_{\text high}$= 0.9. The limitations of finite maximum evolution time $T$ are evident along each row. It is clear that using a larger maximum evolution time $T$ allows the extracted decay rate of a larger $V/\Delta_l$ to fall on the exponential line. While differences along each column highlights the initial state dependence. The outlier points are artifacts of fitting to the wrong regions since this fitting-window selection scheme is sensitive to the oscillation of $y_{\text min}$ and fails to correctly distinguish the quantum Zeno, FVD and oscillation regimes, particular when $\Delta_l$ is small. The plateau at large $V/\Delta_l$ results from fitting to the quantum Zeno regime before FVD sets in. Despite its shortcomings, this fitting procedure still gives the same exponential scaling of the decay rate in quantitative agreement with the aforementioned fitting method.

\section{BCH expansion of AFM order dynamics}\label{SM.sec.BCH}
The Baker-Campbell-Hausdorff (BCH) expansion provides a perturbative series for the short-time Heisenberg evolution. We employ this method to analyze the early-time dynamics of the AFM order parameter,
\begin{equation}
\hat M=\frac{1}{N}\sum_i \varepsilon_i \hat\sigma_i^z,\qquad \varepsilon_i=(-1)^i,\qquad
\langle \hat M\rangle_{\text{N\'eel}}=1,
\label{eq:def-M}
\end{equation}
under the Hamiltonian
\begin{equation}
\hat H=\sum_{i<j} V_{i,j}\hat n_i\hat n_j+\frac{\Omega}{2}\sum_i \hat\sigma_i^x
-\sum_i(\Delta_g-\varepsilon_i\Delta_l)\hat n_i,\qquad
\hat n_i=\frac{1+\hat\sigma_i^z}{2}.
\label{eq:def-H}
\end{equation}
The BCH expansion yields
\begin{equation}
\hat M(t)=e^{i\hat H t}\hat M e^{-i\hat H t}
=\sum_{k\ge 0}\frac{i^k}{k!}\,\hat C_k\,t^k,\qquad
\hat C_k\equiv \mathrm{ad}_{\hat H}^k(\hat M),
\label{eq:zassenhaus}
\end{equation}
with the recursion $\mathrm{ad}_{\hat H}^0(\hat M)=\hat M$ and $\mathrm{ad}_{\hat H}^{k+1}(\hat M)=[\hat H,\mathrm{ad}_{\hat H}^{k}(\hat M)]$.

The first three nested commutators are
\begin{subequations}\label{eq:Ck}
\begin{align}
\hat C_0&=\hat M=\frac{1}{N}\sum_i \varepsilon_i \hat\sigma_i^z,
\label{eq:C0}
\\
\hat C_1&=[\hat H,\hat M]=-\frac{i\Omega}{N}\sum_i \varepsilon_i \hat\sigma_i^y,
\label{eq:C1}
\\
\hat C_2&=[\hat H,\hat C_1]
=\frac{\Omega^2}{N}\sum_i \varepsilon_i \hat\sigma_i^z
+\frac{\Omega}{N}\sum_i\Big[(\varepsilon_i\Delta_g-\Delta_l)-\varepsilon_i\hat{\bar n}_i\Big]\hat\sigma_i^x,\notag \\\quad \hat{\bar n}_i&\equiv\sum_{m\neq i}V_{i,m}\hat n_m.
\label{eq:C2}
\end{align}
\end{subequations}

\subsection{The vanishing odd orders}
For any initial state with real coefficients in the computational basis, all odd-order expansion coefficients vanish. Let $\Theta$ denote the time-reversal operator (complex conjugation in the $\hat\sigma^z$ basis). Since $\Theta \hat H \Theta^{-1}=\hat H$ and $\Theta \hat M \Theta^{-1}=\hat M$, any $|\psi\rangle$ that is real in this basis (e.g., N\'eel or PQG state), $\Theta|\psi\rangle=|\psi\rangle$, satisfies
\begin{equation}
\begin{aligned}
    \langle \hat M(t)\rangle_\psi&=\langle\psi| e^{i\hat H t}\hat M e^{-i\hat H t}|\psi\rangle=\langle\psi|\Theta^2 e^{i\hat H t}\Theta^2\hat M \Theta^2e^{-i\hat H t}\Theta^2|\psi\rangle\\&=\langle\psi| e^{-i\hat H t}\hat M e^{i\hat H t}|\psi\rangle=\langle \hat M(-t)\rangle_\psi,\\
&\Rightarrow\qquad
\langle \hat C_{2m+1}\rangle_\psi=0\ (m=0,1,2,\dots).
\end{aligned}
\label{eq:odd-vanish}
\end{equation}

\subsection{The second order}
For the N\'eel state, $\langle \hat\sigma_i^{x}\rangle=\langle \hat\sigma_i^{y}\rangle=0$, hence
\begin{equation}
\langle \hat C_1\rangle_{\text{N\'eel}}=0,
\qquad
\langle \hat C_2\rangle_{\text{N\'eel}}=\Omega^2\langle \hat M\rangle_{\text{N\'eel}},
\label{eq:C1C2-neel}
\end{equation}
and therefore $\langle \hat M(t)\rangle_{\text{N\'eel}}
=1-\tfrac{\Omega^2}{2}\,t^2+\mathcal O(t^4)$, predicting a decay independent of $\Delta_l$.

In contrast, for the PQG state (the ground state of $\hat H$ at $\Delta_l\to 0^{-}$), we decompose the Hamiltonian as $\hat{H}=\hat{H}_{0} + \hat\delta$:
\begin{equation}
\hat\delta=\Delta_l\hat N,\qquad \hat N\equiv\sum_i\varepsilon_i \hat n_i,\qquad
\hat H_0\equiv \hat H\big|_{\Delta_l=0^-}.
\label{eq:H-split}
\end{equation}
Using $[\hat N,\hat M]=0$ and the eigenstate property of the PQG state, $\langle[\hat H_0,\cdot]\rangle_{\mathrm{PQG}}=0$,
\begin{equation}
\begin{aligned}
\langle \hat C_2\rangle_{\mathrm{PQG}}
&=\Big\langle \big[\hat H_0+\hat\delta,[\hat H_0+\hat\delta,\hat M]\big]\Big\rangle_{\mathrm{PQG}}=\Big\langle \big[\hat\delta,[\hat H_0,\hat M]\big]\Big\rangle_{\mathrm{PQG}}
\\&=\Delta_l\,\Big\langle\big[\hat N,\mathrm{ad}_{\hat H_0}\hat M\big]\Big\rangle_{\mathrm{PQG}}=-\frac{\Delta_l\Omega}{N}\Big\langle\sum_i \hat\sigma_i^x\Big\rangle_{\mathrm{PQG}}.
\label{eq:C2-FV-raw} 
\end{aligned}
\end{equation}
Thus $\langle \hat C_2\rangle_{\mathrm{PQG}}$ is explicitly proportional to the symmetry-breaking field $\Delta_l$.

The value $\langle \hat C_2\rangle_{\mathrm{PQG}}$ can be interpreted as follows: We treat the state $|\text{Néel}\rangle = |\downarrow\uparrow\dots\downarrow\uparrow\rangle$ as the background, on which an $L=1$ bubble at site $i$ can be created by the operator $\hat{\sigma}_i^+$ (for odd $i$) or $\hat{\sigma}_i^-$ (for even $i$). By mapping the spin operators to hard-core boson operators~\cite{pitaevskii2016bose,fetter2003quantum}, the expectation value $\langle\sum_i\hat{\sigma}_i^x\rangle_{\rm{PQG}}$ thus represents the condensate amplitude of the $L=1$ bubble.

After approximating $V_{i,j}$ as a nearest-neighbor  interaction with strength $V$, and assuming $\Delta_g = V$ to map the system onto a transverse-field Ising model (TFIM), and then taking the thermodynamic limit with $N\to \infty$, we can obtain an analytical expression for the ratio of the two expectation values of $\hat{C}_2$, which reads
\begin{equation}\label{SM.eq.ratio_of_C2}
\begin{split}
&\langle\hat{C}_2\rangle_{\text{PQG}} / \langle\hat{C}_2\rangle_{\text{N\'eel}} = \frac{\Delta_l}{\Omega}\lim_{{N\to\infty}} \left(-\frac{1}{N}\right) \Big\langle\sum_i \hat\sigma_i^x\Big\rangle_{\mathrm{PQG,TFIM}} \\
&= \frac{\Delta_l}{\Omega} \times
\,\frac{1}{\pi} \int_{0}^{\pi}
\frac{\,2\Omega - V\cos k\,}{
\sqrt{\,4\Omega^{2} + V^{2} - 4\Omega V \cos k\,}
}\, \mathrm{d}k, \\
&= \frac{\Delta_l}{\Omega} \times \left(\frac{\Omega}{V} + \frac{1}{2}\left(\frac{\Omega}{V}\right)^3 + \mathcal{O}\bigg((\Omega/V)^5\bigg)\right), \\
&\approx\frac{\Delta_l}{V} ~~\text{when}~~ \Omega\ll V, 
\end{split}
\end{equation}
where the second line follows the analytical expression of the transverse magnetization of the TFIM, and the third line follows its low-order expansion in $\Omega/V$. While both $\langle\hat{C}_2\rangle_{\text{PQG}}$ and $\langle\hat{C}_2\rangle_{\text{N\'eel}}$ vanish in the limit $\Omega\to 0$, their ratio approaches a finite value $\frac{\Delta_l}{V}$ rather than unity.

In Fig.~3(a) of the main text, the analytical expression (\ref{SM.eq.ratio_of_C2}) no longer applies because $\Delta_g \neq V$; nevertheless, the ratio $-\frac{\Delta_l}{N \Omega}\langle\sum_i\hat{\sigma}_i^x\rangle_{\rm{PQG}}$ still converges, as $\Omega\to 0$, to a finite value linear in $\Delta_l$, which is $\frac{\Delta_l}{2}\left( \frac{1}{\Delta_g}+\frac{1}{2V-\Delta_g} \right) + \mathcal{O}(\Omega^2)$. This follows from the Hellmann–Feynman theorem combined with second-order perturbation theory for the ground-state energy.

\subsection{The fourth order}
To evaluate the fourth-order commutator,
we define the superoperators
\begin{equation}
\mathcal A\equiv \mathrm{ad}_{\hat H_0},\qquad
\mathcal D\equiv \mathrm{ad}_{\hat N}.
\label{eq:def-superops}
\end{equation}
The operator $\hat C_4$ can be expanded as a $\Delta_l$-polynomial
\begin{equation}
\hat C_4(\Delta_l)=(\mathcal A+\Delta_l\mathcal D)^4\hat M
=\sum_{r=0}^3 (\Delta_l)^{\,r}\,\hat C_4^{(r)},
\label{eq:C4-expansion}
\end{equation}
where (the $r=4$ term vanishes because $\mathcal D\hat M=0$)
\begin{subequations}\label{eq:C4-poly}
\begin{align}
\hat C_4^{(0)}&=\mathcal A^4\hat M,
\label{eq:C4-00}
\\
\hat C_4^{(1)}&=\mathcal D\mathcal A^3\hat M+\mathcal A\mathcal D\mathcal A^2\hat M+\mathcal A^2\mathcal D\mathcal A\hat M,
\label{eq:C4-01}
\\
\hat C_4^{(2)}&=\mathcal D^2\mathcal A^2\hat M+\mathcal D\mathcal A\mathcal D\mathcal A\hat M+\mathcal A\mathcal D^2\mathcal A\hat M,\\
\hat C_4^{(3)} &=\mathcal D^3\mathcal A\hat M.
\label{eq:C4-23}
\end{align}
\end{subequations}

When evaluating the expectation value of $\hat{C}_4$ of the PQG state, only terms with $\mathcal D$ acting outermost survive, which yields
\begin{equation}
\langle \hat C_4\rangle_{\mathrm{PQG}}
=\Delta_l\langle \mathcal D\mathcal A^3\hat M\rangle
+\Delta_l^2\langle \mathcal D^2\mathcal A^2\hat M+\mathcal D\mathcal A\mathcal D\mathcal A\hat M\rangle
+\Delta_l^3\langle \mathcal D^3\mathcal A\hat M\rangle.
\label{eq:C4-FV-keepers}
\end{equation}

For simplicity, we only consider the nearest-neighbor terms $V$ that dominate in $V_{ij}\propto |i-j|^{-6}$. In the small-$\Delta_l$ limit, the first-order $\mathcal{O}(\Delta_l)$ contribution in Eq.~\eqref{eq:C4-01} yields density-dressed single-bubble and pair-exchange blocks,
\begin{equation}
\begin{aligned}
\langle \hat C_4\rangle_{\mathrm{PQG}}^{(1)}=\langle \mathcal D\mathcal A^3\hat M\rangle
&=-\frac{2\Omega}{N}\sum_i V^2\,\Big\langle \hat n_{i-1}\hat\sigma^x_{i}\hat n_{i+1}\Big\rangle_{\mathrm{PQG}}\\
&\quad -\frac{\Omega^2}{N}\sum_i V\,\Big\langle \hat\sigma^x_i\hat\sigma^x_{i+1}+\hat\sigma^y_i\hat\sigma^y_{i+1}\Big\rangle_{\mathrm{PQG}}+\frac{\Omega(2\Delta_g-V)}{N}\sum_i V\,\Big\langle \hat\sigma^x_i(\hat n_{i-1}+\hat n_{i+1})\Big\rangle_{\mathrm{PQG}}
\\
&\quad -\frac{\Omega(\Omega^2+\Delta_g^2)}{N}\sum_i \Big\langle \hat\sigma^x_{i}\Big\rangle_{\mathrm{PQG}} ~,
\end{aligned}
\label{eq:C4-NN-leading}
\end{equation}
which depends on the correlated single-bubble condensate, $\langle \hat n_{i-1}\hat\sigma^x_{i}\hat n_{i+1}\rangle$ and $\langle \hat\sigma^x_i(\hat n_{i-1}+\hat n_{i+1})\rangle$, the $L=1$ bubble condensate density $\langle \hat\sigma^x_{i}\rangle$, and the spin-exchange correlator $\langle \hat\sigma^x_i\hat\sigma^x_{i+1}+\hat\sigma^y_i\hat\sigma^y_{i+1}\rangle$.

When evaluating the expectation value of $\hat{C}_4$ of the Néel state, the lowest term in the $\Delta_l$-polynomial is the zeroth-order, $\mathcal{O}\left((\Delta_l)^0\right)$. Using the definition of the Néel state, Eq.~(\ref{eq:C4-00}) yields:
\begin{equation}
\begin{aligned}
\langle \hat C_4\rangle_{\text{N\'eel}}^{(0)}=\langle \mathcal A^4\hat M\rangle
&=\frac{2\Omega V^2}{N}\sum_i\Big\langle \varepsilon_i\hat n_{i-1}\hat\sigma^z_{i}\hat n_{i+1}\Big\rangle_{\text{N\'eel}}\\
&\quad -\frac{\Omega^2V(2\Delta_g-V)}{N}\sum_i \,\Big\langle \varepsilon_i\hat\sigma^z_i(\hat n_{i-1}+\hat n_{i+1})\Big\rangle_{\text{N\'eel}}
\\
&\quad +\frac{\Omega(\Omega^2+\Delta_g^2)}{N}\sum_i \Big\langle \varepsilon_i\hat\sigma^z_{i}\Big\rangle_{\text{N\'eel}} ~,
\\
&=\Omega^2 \left(2V^2-2\Delta_gV+\Omega^2+\Delta_g^2\right).
\end{aligned}
\label{eq:C4-NN-leading-neel}
\end{equation}
Similar to the case in $\hat{C}_2$, as $\Delta_l\to0$, $\langle \hat C_4\rangle_{\text{N\'eel}}$ approaches a finite value while $\langle \hat C_4\rangle_{\mathrm{PQG}}$ vanishes linearly.

\subsection{Higher orders and practical truncation}
At higher orders, in the small-$\Delta_l$ limit one still finds $\langle\hat C_{2m}\rangle_{\mathrm{PQG}}=\mathcal O(\Delta_l)$, indicating the slow decay of PQG state. At order $\hat C_6$ we encounter length-3 (tri-bubble) NN strings, e.g.
\begin{equation}
\mathcal X_3^{\mathrm{NN\!-\!chain}}=\!\!\sum_{i}\!
\Big\langle \hat\sigma_i^x\hat\sigma_{i+1}^x\hat\sigma_{i+2}^x\Big\rangle,\quad \cdots
\label{eq:X3}
\end{equation}
together with density-dressed single-bubble and density-dressed length-2-bubble strings. While higher BCH orders control progressively longer times, the qualitative decay of $\langle \hat M(t)\rangle$ is already captured by the hierarchy of bubble condensates and their density dressing.

\end{document}